\documentclass[aps,prb,twocolumn,groupedaddress,showpacs]{revtex4-1}

\usepackage{CJK}
\usepackage{graphicx}
\usepackage{amsfonts}
\usepackage{amsmath}
\usepackage{amssymb}
\usepackage{bm}
\begin{document}
\begin{CJK}{GBK}{song}
\title{Long-Range triplet Josephson Current Modulated by the Interface Magnetization Texture}
\author{Hao Meng}
\author{Xiuqiang Wu}
\author{Zhiming Zheng}
\affiliation{National Laboratory of Solid State Microstructures and
Department of Physics, Nanjing University, Nanjing 210093, China}
\date{\today }

\begin{abstract}
We have investigated the Josephson coupling between two s-wave superconductors separated by the ferromagnetic trilayers with noncollinear magnetization. We find that the long-range triplet critical current will oscillate with the strength of the exchange field and the thickness of the interface layer, when the interface magnetizations are orthogonal to the central magnetization. This feature is induced by the spatial oscillations of the spin-triplet state $|$$\uparrow\downarrow\rangle$+$|$$\downarrow\uparrow\rangle$ in the interface layer. Moreover, the critical current can exhibit a characteristic nonmonotonic behavior, when the misalignment angle between interface magnetization and central ferromagnet increases from 0 to $\pi/2$. This peculiar behavior will take place under the condition that the original state of the junction with the parallel magnetizations is the $\pi$ state.\end{abstract}

 \maketitle

  \section {Introduction}
  The interplay between ferromagnetism and superconductivity in hybrid structures has recently attracted considerable attention because of the underlying rich physics and potential applications in spintronics~\cite{1,2}. In a homogeneous ferromagnet (F) adjacent to an s-wave superconductor (S), the Cooper pairs, in which two electrons have opposite spins and momenta ($\textbf{k}$$\uparrow$,$-\textbf{k}$$\downarrow$), can penetrate into the F a short-range. In this case, superconducting correlations between pairs of electrons are induced in both the spin-singlet component $|$$\uparrow\downarrow\rangle$-$|$$\downarrow\uparrow\rangle$ (afterwards, the normalizing factor $1/\sqrt{2}$ is omitted for brevity) and the spin-triplet component $|$$\uparrow\downarrow\rangle$+$|$$\downarrow\uparrow\rangle$ with zero spin projection along the magnetization axis. Accordingly, an opposite-spin Cooper pair $|$$\uparrow\downarrow\rangle$ acquires the total momentum $\emph{Q}$ or $-$$\emph{Q}$ inside the ferromagnet as a response to the exchange splitting 2$\emph{h}_0$ between the spin up and spin down bands~\cite{3}. Here $Q$$\simeq$$2\emph{h}_0/$$\hbar$$v_F$, where $v_F$ is the Fermi velocity. The resulting state is a mixture of singlet component and triplet component with zero total spin projection: ($|$$\uparrow\downarrow\rangle$-$|$$\downarrow\uparrow\rangle$)cos($Q$$\cdot$$R$)+$i$$\cdot$($|$$\uparrow\downarrow\rangle$+$|$$\downarrow\uparrow\rangle$)sin($Q$$\cdot$$R$). These two components oscillate in F with the same period but their phases differ by $\pi/2$.

  In contrast, more interesting the long-range triplet component with parallel electron spins can be induced by the noncollinear magnetic configuration~\cite{4,5,6}. This component penetrates into the ferromagnet over a large distance, which can be on the order of the normal metal coherence length $\xi_N$ in some cases. The reason is that the component with parallel spins is not as sensitive to the exchange field as the opposite-spin component. In order to observe this effect, Houzet \emph{et al.}~\cite{7} suggests measuring the critical current in a Josephson junction through a ferromagnetic trilayers with noncollinear magnetizations. In this geometry, the interface magnetization is in a different direction with the central F layer. This noncollinear magnetic configuration leads to spin-flip scattering processes at the interfaces. It can convert the triplet pairs $|$$\uparrow\downarrow\rangle$+$|$$\downarrow\uparrow\rangle$ into the equal-spin pairs $|$$\uparrow\uparrow\rangle$ or $|$$\downarrow\downarrow\rangle$, and that equal-spin pairs may propagate coherently over long distances into the central F layer. A maximal critical current is obtained when the interface layers have a thickness comparable to the ferromagnetic coherence length $\xi_F$, and the magnetizations in successive layers are orthogonal. Moreover, Alidoust \emph{et al.}~\cite{27} replace the interface homogeneous F layers with the spiral ferromagnetic layers and domainwall ferromagnets. They find a synthesis of 0-$\pi$ oscillations with superimposed rapid oscillations when varying the width of the interface layer. Indeed, many recent experiments have measured a strong enhancement of the long-range Josephson current through a ferromagnetic multilayer with noncollinear magnetizations~\cite{8,9,10}. Theoretically, Hikino \emph{et al.}~\cite{11} shows that the long-range spin current can be driven by the superconducting phase difference in a Josephson junction with double layer ferromagnets. Such spin current carried by the spin-triple Cooper pairs exhibits long-range propagation in the F even when the Josephson charge current is practically zeros.
  
  In this paper, we adopt Blonder, Tinkham, and Klapwijk (BTK) approach~\cite{12} and Bogoliubov's self-consistent field method~\cite{13} to study the critical Josephson current in clean S/$F_1$/$F_2$/$F_3$/S junction with noncollinear magnetization. We propose that the long-range triplet critical current can be modulated by varying the strength of exchange field and thickness of interface ferromagnet, when the interface magnetizations are orthogonal to the central magnetization. This feature is induced by the spatial oscillation of the triplet state $|$$\uparrow\downarrow\rangle$+$|$$\downarrow\uparrow\rangle$ in the interface region. In addition, the critical current manifests a characteristic nonmonotonic behavior, when the misalignment angle between interface magnetization and central ferromagnet changes from 0 to $\pi/2$. This peculiar behavior will appear when the $\pi$ state is the original state of the junction on condition that the ferromagnetic trilayers have the parallel magnetizations.

  \section{Model and formula}
\begin{figure}
  \centering
  \includegraphics[width=3.48in]{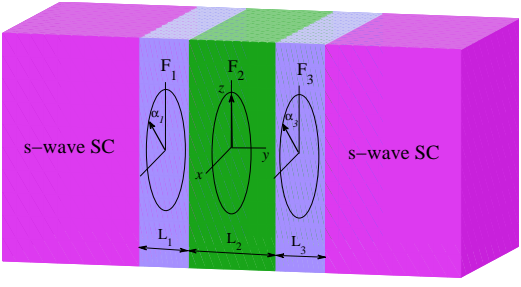} %\\++
  \caption{(color online) Schematic illustration of the S/$F_1$/$F_2$/$F_3$/S junction. The magnetization vectors lie in the \emph{x-z} plane and three arrows indicate the direction of magnetizations in $F_1$, $F_2$ and $F_3$. The phase difference between the two s-wave Ss is $\varphi$$=$$\phi_R$$-$$\phi_L$.}
  \label{fig.1}
  \end{figure}

  We study a clean Josephson junction formed of two s-wave Ss contacted through three F layers with noncollinear magnetization (see fig.~\ref{fig.1}). For convenience, we denote the three ferromagnetic layers sequentially by $F_1$, $F_2$ and $F_3$. The three F layers have the thickness $L_1$, $L_2$, $L_3$, and the exchange field $h_1$, $h_2$, $h_3$. The transport direction is along the \emph{y} axis, and the system is assumed to be infinite in the \emph{x-z} plane.

  The BCS mean-field effective Hamiltonian~\cite{1, 13} is
  \begin{equation}
  \label{Eq1}
  \begin{aligned}
   H_{eff}&=\int{d\vec{r}}\{\sum_{\tilde{\alpha}}\psi^{\dag}_{\tilde{\alpha}}(\vec{r})H_e\psi_{\tilde{\alpha}}(\vec{r})+\frac{1}{2}[\sum_{\tilde{\alpha},\tilde{\beta}}(i\sigma_{y})_{\tilde{\alpha}\tilde{\beta}}\Delta(\vec{r})\\
   &\psi^{\dag}_{\tilde{\alpha}}(\vec{r})\psi^{\dag}_{\tilde{\beta}}(\vec{r})+H.C.]-\sum_{\tilde{\alpha},\tilde{\beta}}\psi^{\dag}_{\tilde{\alpha}}(\vec{r})(\vec{h}\cdot\vec{\sigma})_{\tilde{\alpha}\tilde{\beta}}\psi_{\tilde{\beta}}(\vec{r})\},
  \end{aligned}
  \end{equation}
  where $H_e=-\nabla^{2}/2m-E_F$, $\psi^{\dag}_{\tilde{\alpha}}(\vec{r})$ and $\psi_{\tilde{\alpha}}(\vec{r})$ are creation and annihilation operators with spin $\tilde{\alpha}$. $\vec{\sigma}$ is the Pauli matrices, $m$ is the effective mass of the quasiparticles in both Ss and Fs, and $E_F$ is the Fermi energy. The superconducting gap is denoted by $\Delta(\vec{r})=\Delta(T)[e^{i\phi_{L}}\Theta(-y)+e^{i\phi_{R}}\Theta(y-L_F)]$ with $L_F=L_1+L_2+L_3$. Here, $\Delta(T)$ accounts for the temperature-dependent energy gap that satisfies the BCS relation $\Delta(T)=\Delta_0tanh(1.74\sqrt{T_c/T-1})$ with $T_c$ the critical temperature of the Ss. $\phi_{L(R)}$ is the phase of the left (right) S and $\Theta(y)$ is the unit step function. The exchange field $\vec{h}$ due to the ferromagnetic magnetizations in the $F_i$ (\emph{i}=1, 2, 3) layers is described by $\vec{h}=h_i(sin\alpha_i\hat{e}_{x}+cos\alpha_i\hat{e}_z)$, where $\alpha_i$ is the polar angle of the magnetization, and $\hat{e}_{x(z)}$ is the unit vector along the \emph{x}(\emph{z}) direction.

  By using the Bogoliubov transformation $\psi_{\tilde{\alpha}}(\vec{r})=\sum_{n}[u_{n\tilde{\alpha}}(\vec{r})\hat{\gamma}_{n}+v^{\ast}_{n\tilde{\alpha}}(\vec{r})\hat{\gamma}^{\dag}_{n}]$ and the anticommutation relations of the quasiparticle annihilation and creation operators $\hat{\gamma}_{n}$ and $\hat{\gamma}^{\dag}_{n}$, we have the Bogoliubov-de Gennes (BdG) equation~\cite{1, 13},
  \begin{equation}
  \label{Eq2}
  \breve{H}\Psi{(\vec{r})}=E\Psi{(\vec{r})},
  \end{equation}
  where
  \begin{equation}
  \nonumber
  \label{HMT}
  \breve{H}=
  \begin{pmatrix}
	\hat{H}(\vec{r}) & \hat{\Delta}(\vec{r}) \\
	-\hat{\Delta}^{*}(\vec{r}) & -\hat{H}(\vec{r})
  \end{pmatrix},
  \end{equation}
  $\hat{H}(\vec{r})$=$H_{e}\hat{\textbf{1}}$$-$$h_{z}(\vec{r})\hat{\sigma}_{z}$$-$$h_{x}(\vec{r})\hat{\sigma}_{x}$ and $\hat{\Delta}(\vec{r})$=$i\Delta(\vec{r})\hat{\sigma}_{y}$. Here, $\hat{\textbf{1}}$ is the unity matrix, $\Psi{(\vec{r})}$=$(u_{\uparrow}(\vec{r})$,$u_{\downarrow}(\vec{r})$,$v_{\uparrow}(\vec{r})$,$v_{\downarrow}(\vec{r}))^{T}$ is four-component wave function. The BdG equation can be solved for each superconductor lead and each ferromagnetic layer, respectively. We can have four different incoming quasiparticles, electronlike quasiparticles (ELQs) and holelike quasiparticles (HLQs) with spin up and spin down. For an incident spin-up electron in the left superconductor, the wave function is
  \begin{equation}
  \begin{aligned}
  \Psi^{S}_{L}(y)&=[ue^{i\phi_{L}/2},0,0,ve^{-i\phi_{L}/2}]^{T}e^{ik_{e}y} \\
  &+a_1[ve^{i\phi_{L}/2},0,0,ue^{-i\phi_{L}/2}]^{T}e^{ik_{h}y} \\
  &+b_1[ue^{i\phi_{L}/2},0,0,ve^{-i\phi_{L}/2}]^{T}e^{-ik_{e}y} \\
  &+a'_1[0,-ve^{i\phi_{L}/2},ue^{-i\phi_{L}/2},0]^{T}e^{ik_{h}y} \\
  &+b'_1[0,ue^{i\phi_{L}/2},-ve^{-i\phi_{L}/2},0]^{T}e^{-ik_{e}y}.
  \end{aligned}
  \label{Eq3}
  \end{equation}
  In this process, the coefficients $b_{1}$, $b'_{1}$, $a'_{1}$, and $a_{1}$ describe normal reflection, the normal reflection with spin-flip, novel Andreev reflection, and usual Andreev reflection, respectively. We note that the momentum parallel to the interface is conserved for these processes.

  The corresponding wave function in the right superconductor is
  \begin{equation}
  \begin{aligned}
   \Psi^{S}_{R}(y)&=c_1[ue^{i\phi_{R}/2},0,0,ve^{-i\phi_{R}/2}]^{T}e^{ik_{e}y} \\
  &+d_1[ve^{i\phi_{R}/2},0,0,ue^{-i\phi_{R}/2}]^{T}e^{-ik_{h}y} \\
  &+c'_1[0,ue^{i\phi_{R}/2},-ve^{-i\phi_{R}/2},0]^{T}e^{ik_{e}y} \\
  &+d'_1[0,-ve^{i\phi_{R}/2},ue^{-i\phi_{R}/2},0]^{T}e^{-ik_{h}y},
  \end{aligned}
  \label{Eq4}
  \end{equation}
  where $c_1$, $d_1$, $c'_1$, $d'_1$ are the transmission coefficients, corresponding to the reflection processes described above. The coherence factors are defined as usual, $u=\sqrt{(1+\Omega/E)/2}$, $v=\sqrt{(1-\Omega/E)/2}$ and $\Omega=\sqrt{E^2-\Delta^2}$. $k_{e(h)}=\sqrt{2m[E_F+(-)\Omega]/\hbar^2-k^{2}_{\parallel}}$ are the perpendicular components of the wavevectors with $k_{\parallel}$ as the parallel component.

  The wave function in the $F_i$ layers can be described by transformation matrix~\cite{14} as
  \begin{equation}
  \begin{aligned}
  \Psi^{F}_{i}(y)&=T_i\{[e\cdot{exp(ik^{e\uparrow}_{Fi}y)}+f\cdot{exp(-ik^{e\uparrow}_{Fi}y)}]\hat{e}_{1} \\
  &+[e'\cdot{exp(ik^{e\downarrow}_{Fi}y)}+f'\cdot{exp(-ik^{e\downarrow}_{Fi}y)}]\hat{e}_{2} \\
  &+[g\cdot{exp(-ik^{h\uparrow}_{Fi}y)}+h\cdot{exp(ik^{h\uparrow}_{Fi}y)}]\hat{e}_{3} \\
  &+[g'\cdot{exp(-ik^{h\downarrow}_{Fi}y)}+h'\cdot{exp(ik^{h\downarrow}_{Fi}y)}]\hat{e}_{4}\}.
  \end{aligned}
  \label{Eq5}
  \end{equation}
  Here $\hat{e}_{1}=[1,0,0,0]^{T}$, $\hat{e}_{2}=[0,1,0,0]^{T}$, $\hat{e}_{3}=[0,0,1,0]^{T}$, $\hat{e}_{4}=[0,0,0,1]^{T}$ are basis wave functions, and $k^{e(h)\sigma}_{Fi}=\sqrt{2m[E_F+(-)E+\rho_{\sigma}h_{i}]/\hbar^2-k^{2}_{\parallel}}$ with $\rho_{\uparrow(\downarrow)}$$=$$1(-1)$ are the perpendicular components of wave vectors for ELQs and HLQs. The transformation matrix is defined as $T_{i}=\hat{\textbf{1}}\otimes(cos\frac{\alpha_{i}}{2}\cdot\hat{1}-i\cdot{sin\frac{\alpha_{i}}{2}}\cdot\hat{\sigma}_{y})$. All scattering coefficients can be determined by solving wave functions at the interfaces.
  \begin{equation}
  \Psi^{S}_L(0)=\Psi^{F}_1(0),\partial_{y}[\psi^{F}_{1}-\psi^{S}_{L}]|_{y=0}=2k_FZ_1\psi^{F}_{1}(0);
  \label{Eq6}
  \end{equation}
  \begin{equation}
  \begin{aligned}
  &\Psi^{F}_{i-1}(y_{i-1})=\Psi^{F}_i(y_{i-1}), \\
  &\partial_{y}[\psi^{F}_{i}-\psi^{F}_{i-1}]|_{y=y_{i-1}}=2k_FZ_i\psi^{F}_{i}(y_{i-1});
  \end{aligned}
  \label{Eq7}
  \end{equation}
  \begin{equation}
  \Psi^{S}_{R}(L_F)=\Psi^{F}_3(L_F),\partial_{y}[\psi^{S}_{R}-\psi^{F}_{3}]|_{y=L_F}=2k_FZ_4\psi^{S}_{R}(L_F).
  \label{Eq8}
  \end{equation}
  Where $Z_1$$\sim$${Z_4}$ are dimensionless parameters describing the magnitude of the interfacial resistances. $y_i$ is local coordinate value at the $F_i/F_{i+1}$ interface, and $k_F=\sqrt{2mE_F}$ is the Fermi wave vector. From the boundary conditions, we obtain a system of linear equations that yield the scattering coefficients. With the scattering coefficients at hand, we can use the finite-temperature Green's function formalism~\cite{15,16,17} to calculate dc Josephson current,
  \begin{equation}
  \label{Eq9}
  \begin{aligned}
  &I_{e}(\varphi)=\frac{k_BTe\Delta}{4\hbar}\sum_{k_{\parallel}}\sum_{\omega_{n}}\frac{k_{e}(\omega_{n})+k_{h}(\omega_{n})}{\Omega_{n}}\cdot \\
  &[\frac{a_1(\omega_{n},\varphi)-a_2(\omega_{n},\varphi)}{k_{e}}+\frac{a_3(\omega_{n},\varphi)-a_4(\omega_{n},\varphi)}{k_{h}}],
  \end{aligned}
  \end{equation}
  where $\omega_{n}=\pi{k_B}T(2n+1)$ is the Matsubara frequencies with $n$=0, 1, 2,... and $\Omega_{n}$=$\sqrt{\omega^{2}_{n}+\Delta^{2}(T)}$. $k_e(\omega_{n})$, $k_h(\omega_{n})$, and $a_j(\omega_{n},\varphi)$ with $j$=1, 2, 3, 4 are obtained  from $k_e$, $k_h$, and $a_j$ by analytic continuation $E\rightarrow{i}\omega_{n}$. We find the critical current from $I_c=Max_{\varphi}|I_e(\varphi)|$.

  The above results for Josephson current in the clean limit are obtained in the stepwise approximation for the pair potential. To obtain the time dependent triplet amplitude functions and the local density of the states (LDOS), we solve the BdG equation~(\ref{Eq2}) by Bogoliubov's self-consistent field method~\cite{13,18,19,20}. We put the junction in a one dimensional square potential well with infinitely high walls. Accordingly, the solution in equation~(\ref{Eq2}) can be expanded a set of basis vectors of the stationary states~\cite{21}, $u_{\sigma}(\vec{r})$$=$$\sum_{q}u^{\sigma}_{q}\zeta_{q}(y)$ and $v_{\sigma}(\vec{r})$$=$$\sum_{q}v^{\sigma}_{q}\zeta_{q}(y)$ with $\zeta_{q}(y)$$=$$\sqrt{2/L_d}sin(q{\pi}y/L_d)$. Here $q$ is a positive integer, and $L_d=L_{S1}+L_F+L_{S2}$, $L_{S1}$ and $L_{S2}$ are thickness of left and right superconductors, respectively. The BdG equation~(\ref{Eq2}) is solved iteratively together with the self-consistency condition ~\cite{13}
  \begin{equation}
  \label{Eq10}
  \Delta(y)=\frac{g(y)}{2}\sum_{0\leq{E}\leq\omega_{D}}[u_{\uparrow}(y)v^{*}_{\downarrow}(y)-u_{\downarrow}(y)v^{*}_{\uparrow}(y)]tanh(\frac{E}{2k_BT}),
  \end{equation}
  where $\omega_{D}$ is the Debye cutoff energy and the effective attractive coupling $g(y)$ is a constant in the superconductor and can vanish outside of it. Iterations are not performed until self-consistency is reached, starting from the stepwise approximation for the pair potential. The amplitudes of the triplet pair are defined as follows~\cite{19,20}
  \begin {equation}
  f_{0}(y,t)=\frac{1}{2}\sum_{E>0}[u_{\uparrow}(y)v^{*}_{\downarrow}(y)+u_{\downarrow}(y)v^{*}_{\uparrow}(y)]S(t),
  \label{Eq11}
  \end {equation}
  \begin {equation}
  f_{1}(y,t)=\frac{1}{2}\sum_{E>0}[u_{\uparrow}(y)v^{*}_{\uparrow}(y)-u_{\downarrow}(y)v^{*}_{\downarrow}(y)]S(t),
  \label{Eq12}
  \end {equation}
  where $S(t)=cos(Et)-isin(Et)tanh(E/2k_BT)$. Additionally, the singlet pair amplitude writes as $f_3\equiv\Delta(y)/g(y)$. The above singlet and triplet pair amplitudes are all normalized to the value of the singlet pairing amplitude in a bulk S material. The LDOS is given by
  \begin {equation}
  N(y,\varepsilon)=-\sum_{\tilde{\alpha}}[u^{2}_{\tilde{\alpha}}(y)f'(\varepsilon-E)+v^{2}_{\tilde{\alpha}}(y)f'(\varepsilon+E)],
  \label{Eq13}
  \end {equation}
  where $f'(\varepsilon)=\partial{f}/\partial{\varepsilon}$ is the derivative of the Fermi function. The LDOS is normalized by its value at $\varepsilon=3\Delta_{0}$ beyond which the LDOS is almost constant.
  \section{Results and Discussions}

  In BTK approach, we will use the superconducting gap $\Delta_0$ as the unit of energy. The Fermi energy is $E_F$=$1000\Delta_0$, the interface transparency is $Z_{1-4}$=$0$, and $T/T_c$=$0.1$. We fix the thickness of the $F_2$ layer $k_FL_2$=$200$, the exchange field $h_2/E_F$=$0.6$, and the magnetization direction along the \emph{z} direction that equivalents to $\alpha_2$=0. Interface layers $F_1$ and $F_3$ have the same features, such as misalignment angles $\alpha_1$=$\alpha_3$, thickness $L_1$=$L_3$, and exchange field $h_1$=$h_3$. In self-consistent field method, we consider the low temperature limit and take $k_FL_{S1}$=$k_FL_{S2}$=400, $\omega_{D}/E_F$=0.1, the other parameters are the same as the ones mentioned above.

  \begin{figure}
  \centering
  \includegraphics[width=3.1in]{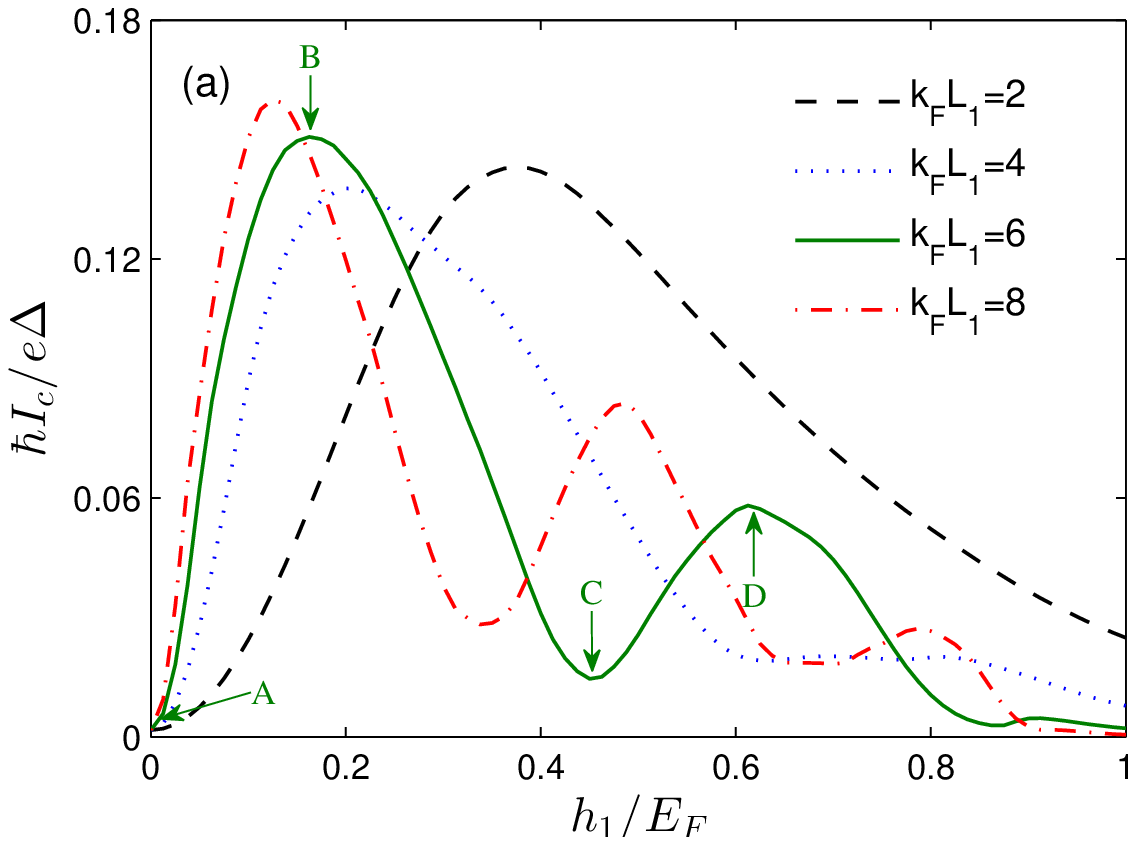} %\\++
  \includegraphics[width=3.1in]{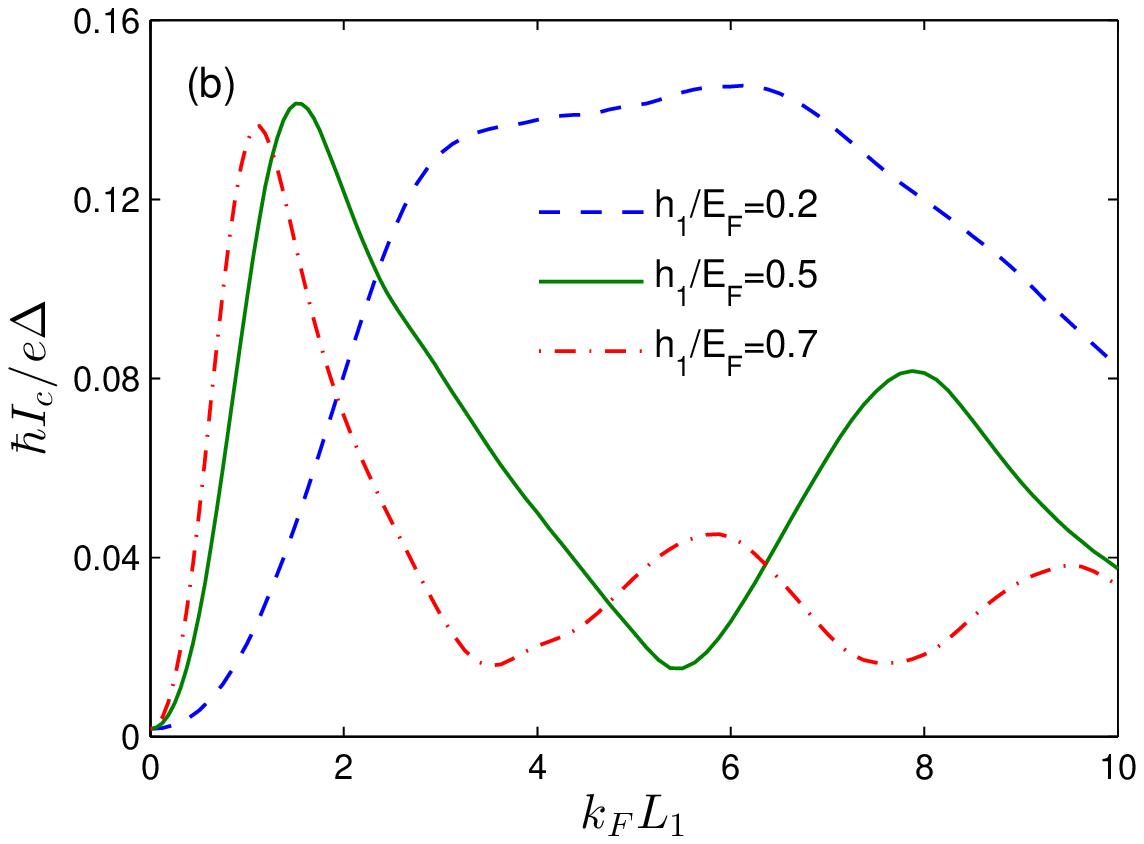} %\\++
  \caption{(Color online) Critical current (a) as a function of the exchange field $h_1$ for different thicknesses $L_1$ and (b) as a function of the thickness $L_1$ for different exchange fields $h_1$ with $\alpha_{1}$=$\pi/2$.}
  \label{fig.2}
  \end{figure}

  Fig.~\ref{fig.2}(a) shows that the critical current is a function of exchange field $h_1$ for the misalignment angles $\alpha_1$=$\pi/2$. We find that, for the thin interface layer $k_FL_1$=2, the critical current $I_c$ only has one peak at exchange field $h_1/E_F$=0.4. With the increase of the thickness $L_1$, the maximum of critical current will gradually move to the left. For the thick interface layer $k_FL_1$=6 and 8, the critical current $I_c$ will display an oscillating behavior as a function of the exchange field $h_1$. The oscillation frequency of critical current $I_c$ can increase with thickness $L_1$. Furthermore, for a fixed exchange field $h_1$, the critical current can display the same oscillatory behavior with the change of the thickness $L_1$ (see fig.~\ref{fig.2}(b)). Especially, for weak exchange field $h_1/E_F$=0.2 (blue dashed line in fig.~\ref{fig.2}(b)), the critical current $I_c$ increases rapidly and reaches a threshold at $k_FL_1$=3, then slowly increases to maximum value at $k_FL_1$=6. Subsequently, the current will decrease at the thicknes $k_FL_1$$>$6. This conclusion is well consistent with the recent work by Khaire \emph{et al.}~\cite{9} and Khasawneh \emph{et al.}~\cite{22}.

  We attribute the above behavior of critical current to two important phenomena at the interface layers~\cite{23,24}: spin mixing and spin-flip scattering process. The exchange splitting in the interface layers $F_1$ will lead to a modulation of the pair amplitude with $Q\cdot{R}$. The resulting state is mixture of spin singlet and triplet state with $S_z$=0. When $F_1$ is magnetized in the same direction as $F_2$ (here, the \emph{z}-axis), the spin singlet and triplet state projected along the \emph{z}-axis, such as ($|$$\uparrow\downarrow\rangle$-$|$$\downarrow\uparrow\rangle$)$_{z}$ and ($|$$\uparrow\downarrow\rangle$+$|$$\downarrow\uparrow\rangle$)$_{z}$, can undergo damped oscillations and penetrate into $F_2$ a short-range. Moreover, if the interface magnetic moment in $F_1$ deviates from the direction of bulk magnetization in $F_2$, this misalignment of the interface moment can lead to spin-flip scattering processes at the interfaces layers. The processes can be described as follows~\cite{3}: as shown in fig.~\ref{fig.1}, the interface layer $F_1$ magnetized in the \emph{x}-direction will generate opposite-spin pairs with respect to the \emph{x}-axis ($|$$\uparrow\downarrow\rangle$+$|$$\downarrow\uparrow\rangle$)$_{x}$. Such a state is equivalent to a combination of equal spin pairs. This pairs projected along the \emph{z}-axis ($|$$\uparrow\uparrow$$\rangle$-$|$$\downarrow\downarrow$$\rangle$)$_z$ can propagate over a long distance into $F_2$. Thus, we can get the following conclusion: long-range triplet Josephson current can modulated by ($|$$\uparrow\downarrow\rangle$+$|$$\downarrow\uparrow\rangle$)$_{x}$ with frequency $Q\cdot{R}$.

  \begin{figure}
  \centering
  \includegraphics[width=3.1in]{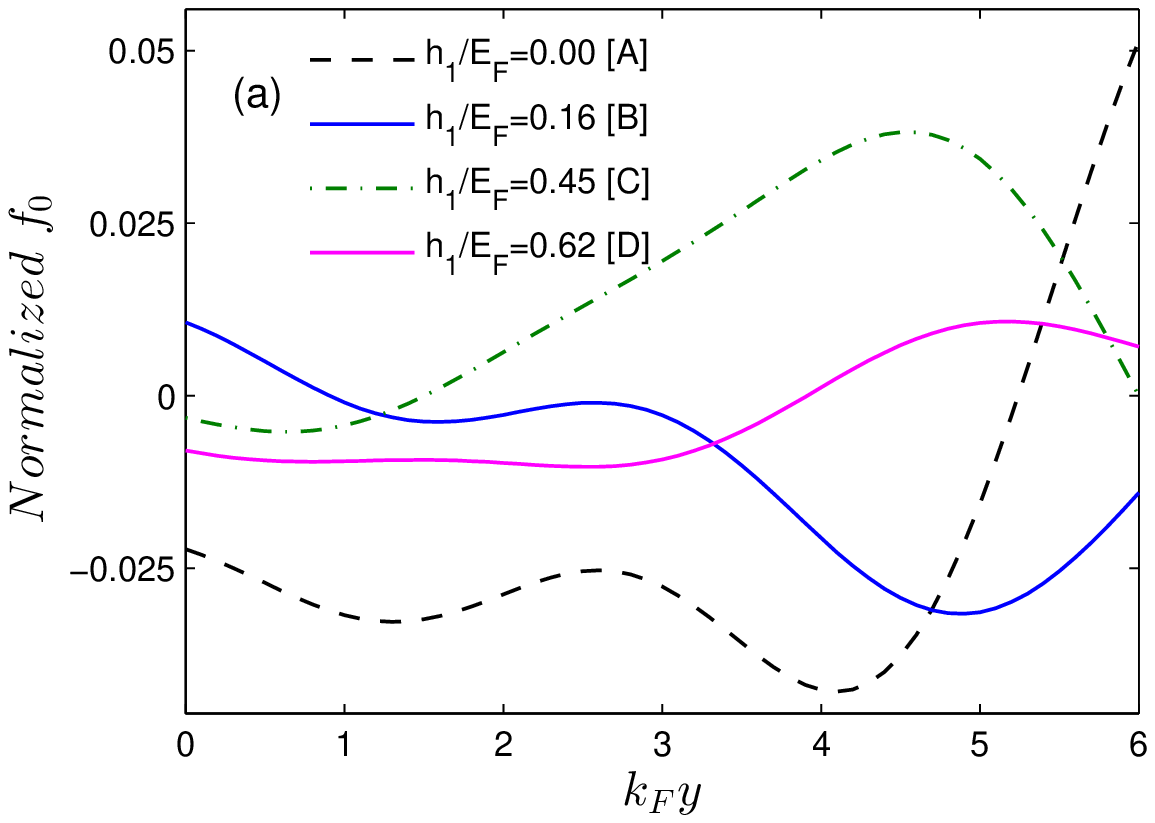} %\\++
  \includegraphics[width=3.1in]{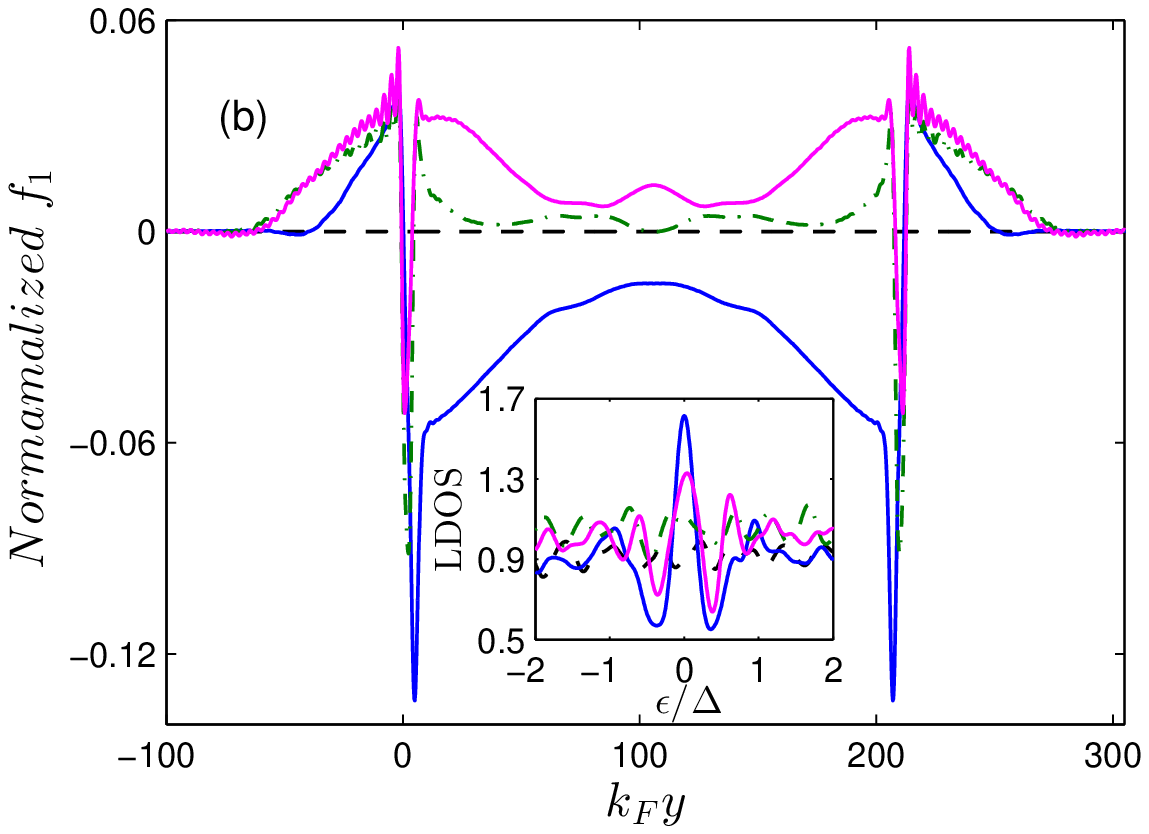} %\\++
  \caption{(Color online) The real part of triplet pair amplitude $f_0$ (a) and $f_1$ (b) plotted as a function of the position $k_Fy$ for different exchange field $h_1$. Here $k_FL_1$=6, $\alpha_{1}$=$\pi/2$ and $t/\omega_{D}$=10. The inset in (b) is the normalized LDOS at $k_Fy$=30, which is calculate at $k_BT=0.001$. Two panels utilize the same legend.}
  \label{fig.3}
  \end{figure}

  Therefore, for thin $R$ (such as $k_FL_1$=2 and $k_FL_1$=4 in fig.~\ref{fig.2}(a)), the relationship $Q\cdot{R}$$<$$\pi$ is always satisfied with the increase of exchange field $h_1$. In this process, the spin-triplet state $f_0$ in $F_1$ will experience only one oscillation, and the critical current has only one peak. For thick $R$ (see $k_FL_1$=6 and $k_FL_1$=8 in fig.~\ref{fig.2}(a)), $Q\cdot{R}$ will undergo many more $0-$$\pi$ transitions, and the $f_0$ in $F_1$ has two such wobbles, or more. Then these $0-$$\pi$ transitions in the phase of $f_0$ will lead to more peaks in the critical current. Similarly, if the $h_1$ is fixed and changes the $L_1$, we can obtain the same conclusions (see fig.~\ref{fig.2}(b)).

  To understanding the above essential physical processes, the spatial distributions of spin-triplet state are calculated in Fig.~\ref{fig.3}. The results show the real part of $f_0$ and $f_1$ for several particular exchange field $h_1/E_F$=0, 0.16, 0.45, and 0.62. These exchange field correspond to the point A, B, C and D in fig.~\ref{fig.2}(a), respectively. For $h_1/E_F$=0, the interface layer $F_1$ is the normal-metal, then the spin-split and spin-flip do not exist. At this moment, the maximum of $f_0$ occurs close to the interface between $F_1$ and $F_2$ at site $k_Fy$=6 (black dashed line in fig.~\ref{fig.3}(a)). Then $f_0$ undergoes a damped oscillations in $F_2$, and can only penetrate a short-range. Meanwhile, $f_1$ does't survive in $F_2$ (black dashed line in fig.~\ref{fig.3}(b)). As a result, the critical current is almost 0. For $h_1/E_F$=0.16, $f_0$ oscillates in $F_1$ and reaches the negative maximum near the interface. In this case, $f_0$ projects along the \emph{x}-axis due to $F_1$ magnetized in the \emph{x}-direction. Such $f_0$ can be converted into the equal-spin state $f_1$, which penetrates over a long distance into $F_2$ (see blue solid line in fig.~\ref{fig.3}(b)). Eventually, the maximal critical current can be obtained. For $h_1/E_F$=0.45, $f_0$ oscillates and reaches to 0 when it gets close to the interface (green dash-dotted line in fig.~\ref{fig.3}(a)). Accordingly, the amplitude of $f_1$ converted from the $f_0$ is small enough, and the critical current can be decreased. For $h_1/E_F$=0.62, $f_0$ retunes to positive amplitude, which can produce a positive spatial distributions of $f_1$. Consequently, the critical current will come back to a larger value. The above changes of the critical current can be also displayed by the LDOS, which is illustrated in the inset of fig.~\ref{fig.3}(b). For $h_1/E_F$=0.16 and 0.62 (correspond to the point B and D in fig.~\ref{fig.2}(a)), the LDOS shows a sharp zero energy conductance peak (ZECP), which represents the long-range $f_1$ is large enough. When $h_1/E_F$=0 and 0.45 (correspond to the point A and C in fig.~\ref{fig.2}(a)), the ZECP will disappear, which illustrates the $f_1$ doesn't exist or is very small.

   \begin{figure}
   \centering
   \includegraphics[width=3.45in]{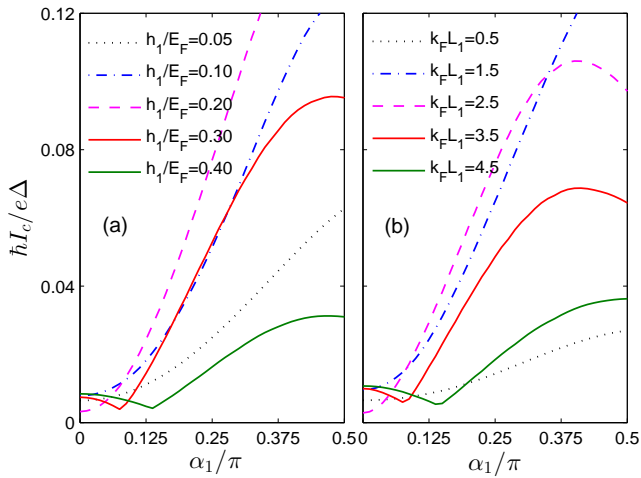} %\\++
   \caption{(Color online) Critical current as a function of misalignment angle $\alpha_{1}$ (a) for different exchange field $h_1$ when thickness $k_FL_1$=6 and (b) for different thickness $L_1$ when exchange field $h_1/E_F$=0.5.}
   \label{fig.4}
   \end{figure}

   In the following, we will discuss how critical current changes with the misalignment angle $\alpha_{1}$. We have displayed the critical current $I_c$ change with the misalignment angle $\alpha_{1}$ for different exchange fields $h_1$ when thickness $k_FL_1$=6 in fig.~\ref{fig.4}(a). For $h_1/E_F$=0.05, 0.1 and 0.2, the current is a monotonous function of the misalignment angle $\alpha_{1}$. However, for $h_1/E_F$=0.3 and 0.4, the current exhibits a nonmonotonic behavior with the increase of the misalignment angle $\alpha_{1}$, where the minimum occurs at some intermediate value and the maximum nearly at $\alpha_{1}$=$\pi/2$. This can be analyzed by the explanation of ref.\cite{25}. The above behavior of critical current is determined by the original phase of junction when the misalignment angle $\alpha_{1}$=0. For a particular exchange field $h^*$, the $0-$$\pi$ transition takes place in the junction just at the given thickness $k_FL_1$=6. Hence, for $h_1$$>$$h^*$ the original state of the junction with $\alpha_{1}$=0 is the $\pi$ state, while for $h_1$$<$$h^*$ it is the 0 state. If one doesn't take absolute value for $I_e(\varphi)$ to define the critical current $I_c$, then $I_c$ is all negative and monotonously decrease when $\alpha_{1}$ changes from 0 to $\pi/2$. In this case, the original state of the junction is the 0 state. However, if the original state is $\pi$ state, the $I_c$ will decrease from a positive quantity to a negative one and reach the negative maximum nearly at the $\alpha_{1}$=$\pi/2$, then the nonmonotonic behavior mentioned beforehand would transform into monotonic, accompanying with the change of sign.

   \begin{figure}
   \centering
   \includegraphics[width=3.1in]{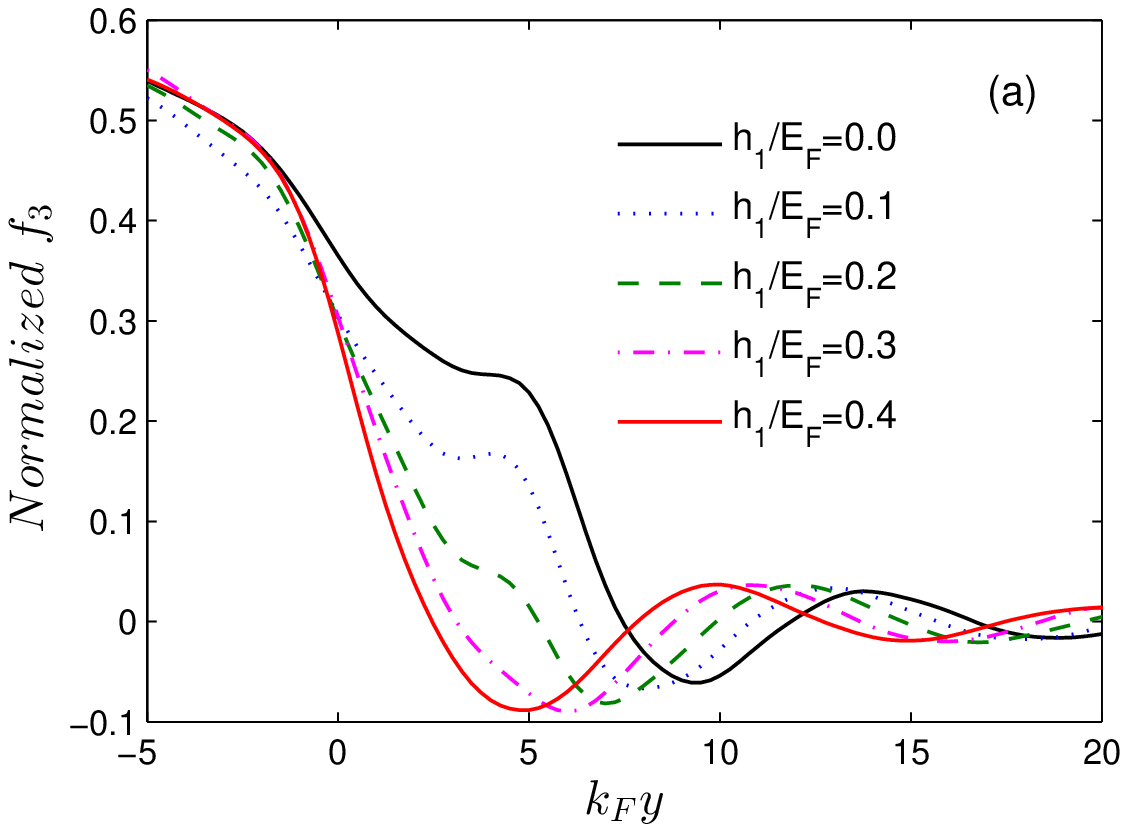} %\\++
   \includegraphics[width=3.1in]{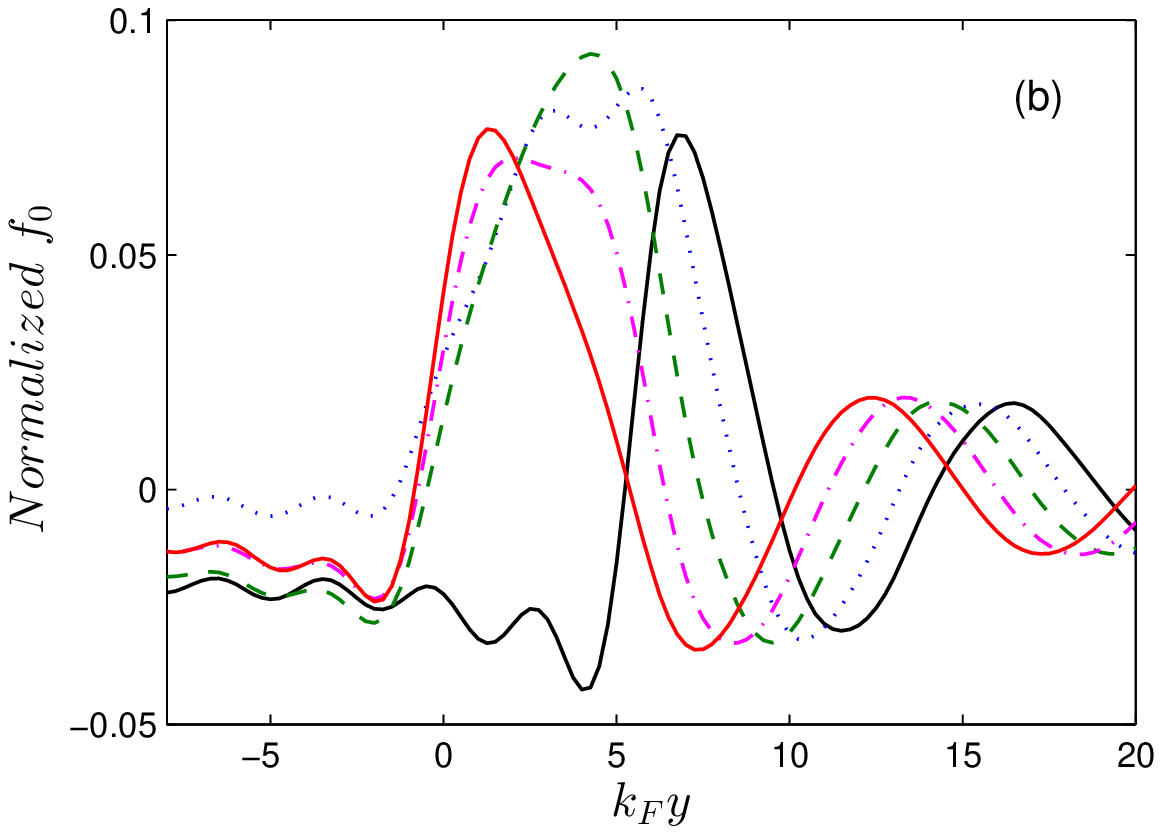} %\\++
   \caption{(Color online) The singlet pair amplitude $f_3$ (a) and the real part of triplet pair amplitude $f_0$ (b) plotted as a function of position $k_Fy$ for different $h_1$. Here $k_FL_1$=6, $\alpha_{1}$=0, and $t/\omega_{D}$=10. Two panels utilize the same legend.}
   \label{fig.5}
   \end{figure}

   To further illustrate this behavior, we plot $f_3$ and the real part of $f_0$ at $\alpha_{1}$=0 and $k_FL_1$=6. In fig.~\ref{fig.5}(a), we display the spatial distributions of $f_3$ for four different values of exchange field $h_1$. With the increase of the exchange field $h_1$, the oscillations of $f_3$ gradually move to the left. At last, their pattern configuration in $F_2$ range would be reversed at $h_1/E_F$$\sim$0.3-0.4. As shown in fig.~\ref{fig.5}(b), $f_0$ has the same behaviors. All of these reversals predict the 0-$\pi$ transition takes place with the increase of exchange field $h_1$. Particularly, the above nonmontonic behavior about the critical current $I_c$ has been observed by Klose \emph{et al.}~\cite{10} and Wang \emph{et al.}~\cite{26}. They apply a large magnetic field to rotate the interface magnetization. As for low fields, a shallow dip can be found in $I_c$. In addition, fig.~\ref{fig.4}(b) shows that the critical current $I_c$ changes with the misalignment angles $\alpha_{1}$ for different thickness $L_1$ at fixed exchange field $h_1$. This behavior is same as fig.~\ref{fig.4}(a).

   \section{Conclusion}
   In this article, we have calculated the Josephson current by BTK approach in clean S$/F_1/F_2/F_3/$S junctions with noncollinear magnetization. Moreove, to demonstrate changes of the critical current, we also study the singlet and triplet pair amplitude and the local density of states in ferromagnets by using Bogoliubov's self-consistent method. We have shown that the dependence long-range triplet critical current with the strength of exchange field and thickness of the interface layers shows an oscillatory behavior, when the interface magnetizations are orthogonal to the central magnetization. This change of the critical current is induced by the oscillations of ($|$$\uparrow\downarrow$$\rangle$+$|$$\downarrow\uparrow$$\rangle$)$_x$ with frequency $Q\cdot{R}$. In addition, when the orientations of the magnetizations in interface layers rotate from \emph{z}-axis to \emph{x}-axis, the critical current always exhibits a nonmonotic behavior. We have discussed the condition for this effect to be observed under the condition that the original state of the junction with the parallel magnetizations is the $\pi$ state.

   \acknowledgments
    The authors are grateful to Prof. Shi-Ping Zhou and Dr. Feng Mei for useful discussions. This work is supported by the State Key Program for Basic Research of China under Grants No. 2011CB922103 and No. 2010CB923400, and the National Natural Science Foundation of China under Grants No. 11174125 and No. 11074109.

    \end{CJK}
    \end{document}